\documentclass{article}

\usepackage{graphicx} % Include figure files
\usepackage{amsmath}
\usepackage{amssymb}
\usepackage{bm} % bold math
\usepackage{cite}

\begin{document}

%%% ---------------------------------------------------------

% This is the title block for ARXIV.

\begin{center}
{\huge\textbf{The Lambert W Function, \\
   Laguerre Polynomials,\\
   and the Zeros of the QCD \\
   Partition Function}}\\
--------------------------------------------------------------------\\
Ken Roberts\footnote{Physics and Astronomy Department, 
Western University, London, Canada, krobe8@uwo.ca},
S. R. Valluri\footnote{Physics and Astronomy,
and Applied Mathematics Departments,
Western University, London, Canada;
King's University College, London, Canada,
valluri@uwo.ca, vallurisr@gmail.com}\\
July 3, 2013
\end{center}

\begin{abstract}
We study solutions of a transcendental equation for the
complex chemical potential at which a random-matrix QCD 
model can undergo a phase transition at zero mass.  
An explicit solution is obtained in terms of the 
Lambert $W$ function.
We also provide a closed form expression for a QCD
random matrix model partition function, as a sum of 
Laguerre polynomials, for complex chemical potential
and non-zero mass.
\end{abstract}

% \pacs{12.38.Gc, 05.70.Fh}

% \keywords{Lambert W function, Laguerre polynomial,
 % complex chemical potential,
 %  QCD partition function, Yang-Lee zeros}

% \maketitle

% For Arxiv, all the Latex code is in this one file.

\section{Introduction}
\label{sect-intro}

The study of zeros of the partition function 
is a classic problem of mathematical physics
and an essential tool for the investigation 
of phase transitions (see \cite{010-Huang-1987}, section 9.3).
Yang and Lee \cite{020-Yang-Lee-1952, 030-Lee-Yang-1952}
showed that, subject to reasonable assumptions,
the thermodynamic limit of the
Helmholtz free energy density is an analytic function 
of the fugacity in any region of the complex fugacity plane
(or the complex chemical potential plane)
in which the partition function has no zeros.

In lattice quantum chromodynamics (QCD), 
the connection between phase transitions 
and the complex chemical potential has received much study;
see Aarts \cite{040-Aarts-2013, 050-Aarts-James-2012} 
for recent status reports.
There are difficult issues arising in the simulation
of the partition function of lattice QCD for complex
chemical potential.

%% see ref 10 of Aarts-2013, ie Osborn-Splittorff-Verbaarschot paper.

Halasz, Jackson, and Verbaarschot [hereafter HJV]
considered \cite{060-Halasz-1997} a random matrix partition function 
with the global symmetries of the QCD partition function,
and obtained interesting results.
They utilized both numerical and analytical methods
to determine the Yang-Lee zeros of the partition
function $Z_N(m,\mu)$ for the system considered.
Here $m$ is the mass parameter, which
is allowed to be complex-valued,
$\mu$ is the complex chemical potential,
and $N$ is a ``size-of-system" parameter which
is analogous to volume.
Their study included explorations in each of the
complex mass plane and the complex chemical
potential plane.

HJV obtained two results which
are the focus of our present work.  
These are found in the original HJV paper
\cite{060-Halasz-1997} as equations (8)
and (10), respectively.

Equation (8) of the HJV paper is a transcendental
equation, expressing the conditions on $\mu$ for a phase
transition when the mass $m$ is zero in their model.
This equation can be solved explicitly using the
Lambert $W$ function.
We present the details of that solution
in section \ref{sect-eqn8}.
Consideration of the multiple branches of the
Lambert $W$ function exhibits some additional
solutions to those obtained in the HJV paper.

Equation (10) of the HJV paper is the
QCD partition function for their random matrix model.
This can be written as a sum of Laguerre polynomials.
We present the details in section \ref{sect-eqn10}.

There are two advantages which may be obtained
from the expression of formulas
which arise in a model of a particular problem,
in terms of well-known functions. 
Firstly, we can use our knowledge of those
standard functions for improved understanding of the
behavior of the formulas in the problem at hand, 
and perhaps gain insight into the 
underlying physical and analytical relationships.
Secondly, in conjunction with highly efficient software
for the calculation of standard functions, it may be
possible to speed up the computational work for the
problem model.

This paper is a particular topic within a more
broad-focus study of the concepts of complex chemical
potential and complex fugacity.
We are interested in the manner in which extending
the domain of analysis for chemical potential and
fugacity into the complex plane, can provide further
insight into physical models.
Our inspiration has been the papers of M. Howard Lee
\cite{070-MHLee-1995, 080-MHLee-1997} regarding
chemical potential and the use of polylogarithms
in statistical mechanics.
Here, we direct our attention
to equations (8) and (10) of the HJV paper
\cite{060-Halasz-1997}.

\section{Lambert W Solution of HJV Eqn (8)}
\label{sect-eqn8}

The random matrix model which was explored by HJV
involves replacing the matrix elements of the Dirac
operator in the continuum QCD partition function,
with Gaussian distributed random variables consistent
with the global partition function.
See Verbaarschot and Wettig \cite{090-Verbaarschot-2000}
for a review of the technique.
HJV in \cite{060-Halasz-1997} consider a single flavor of quarks.

If $m = 0$, then in the thermodynamic limit
the points $\mu$ at which the phase change occurs
are determined by the solutions to the transcendental
equation
\begin{equation}
\label{hjveqn8}
  \mathtt{Re} 
     \big[ 
           \mu^2 + \mathtt{log} (\mu^2)
     \big]
   = -1
\end{equation}
which appears as equation (8) in \cite{060-Halasz-1997}.
The solutions of (\ref{hjveqn8}) are
\begin{equation}
\label{hjvsoln}
  \mu = W(z)^{1/2}
\end{equation}
where $W$ is the multi-branch Lambert $W$ function,
and $z$ is any point on the circle of radius
$1/e$ about the origin.

Before giving the proof, we provide some 
background.
The Lambert $W$ function
$w = W(z)$
is the multi-branch analytic
function which satisfies
\begin{equation}
\label{lamweqn}
  w e^w = z
\end{equation}
Here $w$ and $z$ are complex.
$W_k$ denotes the $k$-th branch of $W$.
For details and history of the Lambert $W$ function,
see Corless, et al \cite{100-Corless-1996}.
For an example of its use in quantum statistics, see
Valluri, et al \cite{110-Valluri-2009}.
Of possible relevance in QCD contexts,
the Lambert $W$ function is also defined
for matrix arguments; 
see Higham \cite{120-Higham-2008}, page 51.

Now we turn to the proof that equation (\ref{hjvsoln})
is the solution of (\ref{hjveqn8}).
Set $w = \mu^2$.
Then
\begin{equation}
\label{weqn1}
  e^{\mu^2 + \mathtt{log}(\mu^2)} 
  = e^{w + \mathtt{log}(w)} = w e^w
\end{equation}
Let $z = w e^w$ so that $w = W(z)$, where $W()$ denotes
the multi-branch Lambert W function, and $w$ is any
of the possible function values.
Suppose that (\ref{hjveqn8}) holds.  That is,
\begin{equation}
  \mu^2 + \mathtt{log}(\mu^2) = -1 + \mathtt{i} y
\end{equation}
for some arbitrary real value $y$.  Then
\begin{equation}
\label{weqn2}
  e^{\mu^2 + \mathtt{log}(\mu^2)} 
    = w e^w
    = e^{-1+\mathtt{i} y}
    = \big( {1 \over e} \big) e^{\mathtt{i} y}
    = \big( {1 \over e} \big) v
\end{equation}
where $v$ is an arbitrary point on the unit circle.

Let $C$ denote the circle of radius ${1/e}$ about
the origin.
The solution set of (\ref{hjveqn8}) is thus $\mu^2$
in $W(C)$, the image of $C$ under the Lambert $W$ function.
Taking the square root (another multivalued function),
the solution set of equation (\ref{hjveqn8}) is seen to be
$\mathtt{sqrt}(W(C))$.  That establishes (\ref{hjvsoln}).

\begin{figure}
\centering
\includegraphics[width=75mm,height=75mm]
   {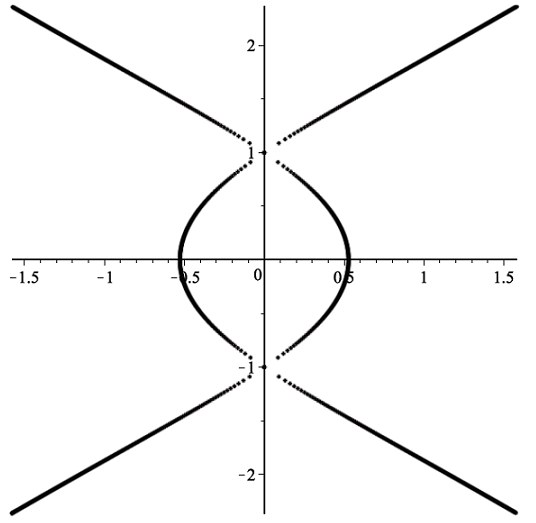}
\caption{\label{fig1}Solutions of equation (\ref{hjveqn8}).}
\end{figure}

What does the solution set $S = \mathtt{sqrt}(W(C))$ look like?
One has to consider the various branches of the
Lambert $W$ function, as well as the two branches of
the square root function.
We have graphed the solution set with
two standard mathematical software packages
(Maple and Mathematica) with identical results.
In figure \ref{fig1} we show
the solution set points for the principal branch $W_0$,
comprising the two central arcs which meet at
$\mu = \mathbf{i}$ and $\mu = -\mathbf{i}$;
the solution set arising from the branch $W_1$,
which is the lines going upward-right and downward-left;
and the solution set arising from the branch $W_{-1}$,
which is the lines going upward-left and downward-right.
The solutions for other branches $W_k$ continue those
lines further.

Compare with figure 2 in \cite{060-Halasz-1997}, the
top portion for $m = 0$: Only the solutions arising
from the principal branch $W_0$ appear in the
figure in \cite{060-Halasz-1997}.
Thus, understanding that equation (\ref{hjveqn8}) has
a solution in terms of the Lambert $W$ function leads
to additional solutions $\mu$ which can be considered
in the larger model context as the possible locations of 
phase transitions.

The real valued solutions of (\ref{hjveqn8}) are readily
calculated with the Lambert $W$ function evaluator 
provided by standard mathematical software.  They are
\begin{equation}
 \mu = \mathtt{sqrt} \Big[ W_0 \Big( {1 \over e} \Big) \Big] 
  = \pm \,\, 0.5276973970
\end{equation}

We note that equation (\ref{hjveqn8}) can also be viewed
in terms of the polylogarithm function $\mathtt{Li}_s$
in this case of order $s=1$,
since $\mathtt{Li}_1(z) = - \mathtt{log}(1-z)$.
There is a potential for generalization to other orders
of polylogs.
We believe that the inherent structure of the 
polylogarithm function will be a useful tool for the
study of properties of partition functions.
See \cite{130-NIST-2010} or 
\cite{140-Lewin-1981} for background 
regarding polylogs, and, for instance, 
\cite{080-MHLee-1997} for an application.

\section{Laguerre Polynomial Rewrite of HJV Eqn (10)}
\label{sect-eqn10}

HJV rewrite the partition function
explicitly as a polynomial in powers of
$\mu^2$ and $m^2$, to obtain
their equation (10), which is
\begin{equation}
\label{hjveqn10}
 Z_N(m,\mu) = {{\pi N!}\over{N^{N+1}}} \,
  \sum_{k=0}^N{ \sum_{j=0}^{N-k}{
  {{(N m^2)^k}\over{(k!)^2}}
  {{(-N\mu^2)^j} \over {j!}} 
  {{(N-j)!} \over {(N-j-k)!}} }}
\end{equation}
For simplicity, let $u = Nm^2$, $v = N\mu^2$,
let $A$ denote the coefficient 
${{\pi N!}\over{N^{N+1}}}$,
and adopt the convention that $j$ and $k$ are
always non-negative.
Then (\ref{hjveqn10}) becomes
\begin{equation}
 Z_N(m,\mu) =A \,
  \sum_{k=0}^N{ \,\, \sum_{j \le N-k}{ \,\,
  {{u^k}\over{k!}} \,\,
  {{(-v)^j} \over {j!}} \,\,
  {{N-j}  \choose {N-j-k}} }}
\end{equation}
Reversing the order of summation and rearranging,
\begin{equation}
 Z_N(m,\mu) =A \,
  \sum_{j=0}^N{ \,\, B_j}
\end{equation}
where
\begin{equation}
 B_j = 
  {{(-v)^j} \over {j!}} \,\,
  \sum_{k \le N-j}{ \,\,
  {{u^k}\over{k!}} \,\,
  {{N-j}  \choose {N-j-k}} }
\end{equation}

The Laguerre polynomials $L_r(x)$, for any non-negative
integer $r$, can be defined in a variety of ways.
See \cite{130-NIST-2010}
for background.
One expression for the Laguerre polynomials
(see, for instance, 
\cite{150-Prudnikov-1986}, page 614, formula 21)
is
\begin{equation}
  L_r(x) = \sum_{s=0}^{r}
  {
    {r \choose s} \,\, {{(-1)^s} \over {s!}} \,\, x^s
  }
\end{equation}
Hence $B_j$ is seen to be 
\begin{equation}
 B_j = 
  {{(-v)^j} \over {j!}} \,\, L_{N-j}(-u)
\end{equation}

We obtain a representation of the partition
function as a sum of Laguerre polynomials,
\begin{equation}
\label{partlagu}
 Z_N(m,\mu) = {{\pi N!}\over{N^{N+1}}} \,
  \sum_{j=0}^N{ \,\, {{(-N\mu^2)^j} \over {j!}} \,\, L_{N-j}(-Nm^2)}
\end{equation}
In particular, for the case of zero chemical potential,
since only the $j=0$ term remains,
$Z_N(m,0)$ is explicitly given by a single Laguerre polynomial
\begin{equation}
 Z_N(m,0) = {{\pi N!}\over{N^{N+1}}} \,  L_{N}(-Nm^2)
\end{equation}
These representations offer the possibility of 
speeding up the computation
of the partition function by using standard mathematical
software.
In this regard, refer to \cite{160-Borwein-2008} on
asymptotic behavior of Laguerre polynomials,
and to \cite{170-Zarzo-1994, 180-Arriola-1991} on
the density of zeros of Laguerre polynomials.

The partition function (\ref{partlagu}) can be expressed
in terms of a confluent hypergeometric function or a
cylindrical Bessel function.
The density of zeros of this partition function can be
studied in terms of the zeros of Bessel functions
\cite{190-Chishtie-2008}.

%% JWKB mentioned see Zarzo ref 7, Froman

%% 0F1 conf hyper geom fcn = Bessel Jn if n goes to infty

\section{Conclusion}
\label{sect-conclusion}

In conclusion, we express the solutions
of the transcendental equation (8) of 
\cite{060-Halasz-1997} for complex $\mu$ 
in terms of the Lambert $W$ function.
We provide an expression
of the QCD random matrix partition function,
equation (10) of \cite{060-Halasz-1997},
as a Laguerre polynomial in $m^2$ 
for zero chemical potential, 
and as a sum of Laguerre polynomials 
for non-zero $\mu$. 
We hope that the analytic results found will be useful 
in understanding QCD random matrix models,
and will result in higher accuracy and a 
reduction of computer time in further studies.

We thank John Drozd for performing the Mathematica 
calculations.
SRV thanks King's University College for their generous support.

\end{document}